\documentclass[%
reprint,
prx,
superscriptaddress,
twocolumn,
amsmath,amssymb,
aps
]{revtex4-2}
\usepackage[english]{babel}
\usepackage[T1]{fontenc}
\usepackage{amsmath,amsthm,amssymb,amsfonts,graphicx,makeidx,hyphenat}
\usepackage{xcolor}
\usepackage{dcolumn}
\usepackage{bm}
\usepackage{url}
\usepackage{braket}
\usepackage{mathrsfs}
\usepackage{dsfont}
\def \tr {\text{Tr}}

\usepackage[colorlinks,colorlinks=true,linkcolor=blue,citecolor=red,linktocpage=true]{hyperref}

\begin{document}

\title{Conditional fluctuation theorems and entropy production for monitored\\ quantum systems under imperfect detection}

\author{Mar Ferri-Cortés}
\thanks{These two authors contributed equally to this work.}
\affiliation{Institute for Cross-Disciplinary Physics and Complex Systems IFISC (UIB-CSIC), \\ E-07122 Palma de Mallorca, Spain}
\affiliation{Departamento de F\'isica Aplicada, Universidad de Alicante, 03690 San Vicente del Raspeig, Spain}

\author{José A. Almanza-Marrero}%
\thanks{These two authors contributed equally to this work.}
\author{Rosa L\'opez}
\author{Roberta Zambrini}
 \author{Gonzalo Manzano}
\email{gonzalo.manzano@ifisc.uib-csic.es}
\affiliation{Institute for Cross-Disciplinary Physics and Complex Systems IFISC (UIB-CSIC), \\ E-07122 Palma de Mallorca, Spain}


\begin{abstract}
The thermodynamic behavior of Markovian open quantum systems can be described at the level of fluctuations by using continuous monitoring approaches. However, practical applications require assessing imperfect detection schemes, where the definition of main thermodynamic quantities becomes subtle and universal fluctuation relations are unknown. Here, we fill this gap by deriving a universal fluctuation relation that links thermodynamic entropy production and information-theoretical irreversibility along single trajectories in inefficient monitoring setups. This relation provides as a corollary an irreversibility estimator of dissipation using imperfect detection records that lower bounds the underlying entropy production at the level of visible trajectories. We illustrate our findings with a driven-dissipative two-level system following quantum jump trajectories and discuss the experimental applicability of our results for thermodynamic inference.
\end{abstract}

\maketitle

\section{Introduction} 
Open quantum systems are subjected to interactions with the environment that can notably alter their properties and evolution from both informational and thermodynamical points of view~\cite{Breuer02,Wiseman2009}. The noisy character of  energy and matter exchanges can be accessed through the application of indirect quantum measurement schemes that monitor system observables or fluxes continuously in time~\cite{Carmichael1993,Belavkin89,Dalibard1992,Dum92,Molmer93}. The output records reveal information about quantum and thermal fluctuations beyond the density matrix approach and allows the extension of concepts in nonequilibrium stochastic thermodynamics to the quantum realm~\cite{Horowitz2012,Hekking2013,Horowitz2013,Suomela2015,Manzano2015,Liu2016,Alonso2016,Auffeves2017,Manzano2018b,Gherardini2018b,Mohammady2020,DiStefano2018,Belenchia2020,Rossi2020,Miller2021,Carollo2021} (for a recent review see Ref.~\cite{Manzano22}). Within this context, a crucial result encoding the fluctuating behavior of thermodynamic quantities is the fluctuation theorem (FT),
\begin{eqnarray} \label{eq:FT0}
    \langle e^{-S_\mathrm{tot}} \rangle = 1,
\end{eqnarray}
where $S_\mathrm{tot}$ is the total stochastic entropy production in the ideal monitored setting [see Eq.(\ref{eq:Stot}) below] and $\langle \cdot \rangle$ denotes the ensemble average over records~\cite{Seifert2012,Manzano22}. As corollary, the FT above implies the second-law inequality $\langle S_\mathrm{tot} \rangle \geq 0$. {The entropy production is a fundamental quantity in nonequilibrium thermodynamics~\cite{Spohn1978,Esposito2010,Deffner2011,Landi21}. It both captures the net energy dissipation in a thermodynamic process in the form of heat (e.g. work dissipated as heat in a driven-dissipative system) and at the same time provides a quantitative measure of irreversibility, that is, of the asymmetry of the thermodynamic process under time-reversal~\cite{Parrondo09,Batalhao2015,Rubino21}.} Moreover, beyond the FT~\eqref{eq:FT0}, quantum versions of other related universal results have been also obtained in Markovian open quantum systems, such as the thermodynamic and kinetic uncertainty relations~\cite{Carollo2019,Hasegawa2020,Hasegawa2021,VanVu22,VanVu23}, or the martingale theory for entropy production~\cite{Manzano2019,Manzano2021}. Altogether they provide new insights on the operation of quantum heat engines~\cite{Campisi2015,Liu2020,Brandner2020}, feedback control scenarios~\cite{Strasberg2013,Gong2016,Murashita2017,Mitchison21,Sagawa22} or information erasure~\cite{Miller2020,Saito22}.

\begin{figure*}[t]
    \centering
    \includegraphics[width=0.95\linewidth]{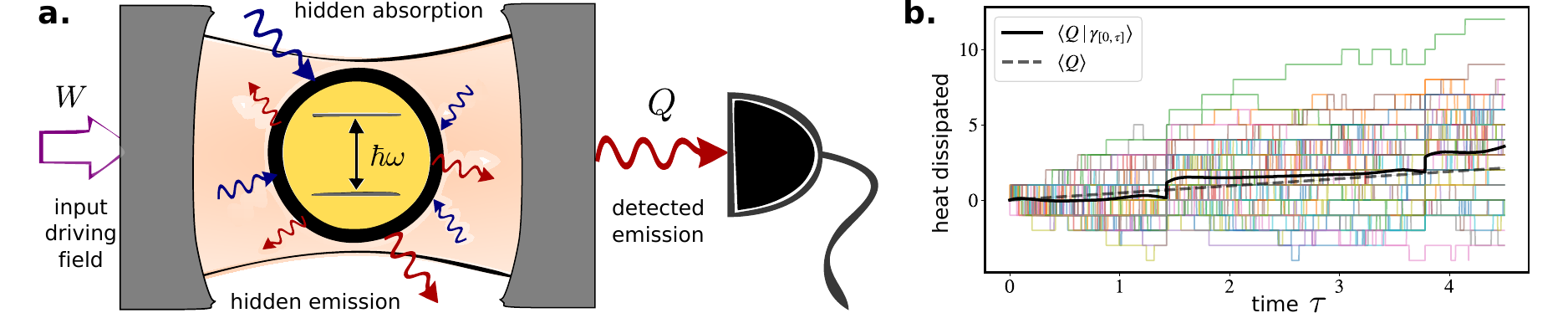}
    \caption{a) Sketch illustrating imperfect detection for a driven two-level atom in an optical cavity. b) Samples of the stochastic heat dissipated along ideal trajectories $\Gamma_{[0,\tau]}$ as a function of time (multiple color lines) conditioned on given set of (two) detected events, together with their conditional average $\langle Q | \gamma_{[0,\tau]} \rangle$ (black thick line) for that visible trajetory $\gamma_{[0,\tau]}$, and the standard average $\langle Q \rangle$ predicted by the Lindblad master equation (grey dashed line).}
    \vspace{-0.3cm}
    \label{fig:sketch}
\end{figure*}

Although any Markovian open quantum system modeled by the Lindblad master equation could be interpreted as monitored by its environment \cite{Wiseman2009}, on the practical side, the actual accessible measurements in the laboratory are always imperfect, unavoidably leading to information losses~\cite{Warszawski02,Oxtoby05,Murch2013,Huard2016,Naghiloo2018,Rossi2019,Minev2019} (see Fig.~\ref{fig:sketch} for an illustrative sketch). Despite the practical importance of such circumstances, little is still known about how thermodynamic quantities can be estimated by using only partially accessible information from the measurement records or what universal properties their fluctuations should verify if any~\cite{Borrelli15,Viisanen15,Elouard2017b,Harrington2019,Naghiloo2020,Kewming22}. These are important questions in order to assess the energetic costs of quantum computation and other quantum technologies~\cite{Auffeves22}, as well as for designing and interpreting experiments on quantum thermodynamics in driven dissipative systems, or in setups with different thermal contacts. 

In this work we extend the thermodynamics of quantum monitored system to partial and imperfect detection {by establishing universal nonequilibrium properties of the total entropy production in that context. In particular, even if under imperfect monitoring the equivalence of the notions of dissipation and irreversibility does not hold anymore, we show how these concepts continue to have a deep relation at the statistical level of fluctuations.} In order to do that, we consider a generic stochastic information-theoretical irreversibility estimator $\Sigma(\gamma_{[0,t]})$ based on the probability of measurement records $\gamma_{[0,t]}$ obtained under imperfect monitoring of the open system up to time $t$, and show how it is related to the thermodynamic entropy production by means of a new conditional fluctuation relation:
\begin{equation} \label{eq:FT}
    e^{-\Sigma} = \langle e^{-S_\mathrm{tot}} | \gamma_{[0,t]} \rangle,
\end{equation}
where $\Sigma$ will be specified below [see Eq.(\ref{eq:Sigma})], and $\langle \cdot | \gamma_{[0,t]} \rangle$ is the conditional average for a given fixed imperfect record. {In contrast to the FT in Eq.~\eqref{eq:FT0}, which contains an average over all trajectories, the above fluctuation theorem~\eqref{eq:FT} contains an average in the r.h.s. only over the hidden information, and is valid for every visible trajectory of the system $\gamma_{[0,t]}$ obtained under imperfect monitoring.} It also implies, by means of Jensen's inequality for conditional expectations, the inequality:
\begin{eqnarray} \label{eq:bound}
    \Sigma(\gamma_{[0,t]}) \leq \langle S_\mathrm{tot} | \gamma_{[0,t]} \rangle,
\end{eqnarray}
which allows us to interpret $\Sigma(\gamma_{[0,t]})$ as a (trajectory) irreversibility estimator lower-bounding stochastic entropy production when the record $\gamma_{[0,t]}$ is observed. Taking the ensemble average of the above equation over many records, we obtain $\langle \Sigma \rangle \leq \langle S_\mathrm{tot} \rangle$, providing a lower bound on the average dissipation of the monitored process, but also $\langle \Sigma^k \rangle \leq \langle S_\mathrm{tot}^k \rangle$ for all even powers ($k= 2, 4, \dots$). 

In the following sections, we provide details on how Eqs.~\eqref{eq:FT} and \eqref{eq:bound} are derived, together with related results refining our understanding of the statistical properties of the link of $\Sigma$ with entropy production [Eqs.~\eqref{eq:min} and \eqref{eq:boundsq}], and discuss their physical interpretation and implications. {Extra information and detailed proofs are given in the Appendices. We illustrate our results for the case of an imperfectly monitored driven-dissipative two-level system in Sec.~\ref{sec:5}. In Sec.~\ref{sec:6} we discuss the experimental implementation of our results and the applicability of $\Sigma$ for thermodynamic inference. Finally, we conclude with some final remarks and outlook of future research directions in Sec.~\ref{sec:7}.}

\section{Framework} \label{sec:2}
We consider Markovian open quantum systems weakly interacting with one or several thermal reservoirs, whose evolution is described by the Lindblad master equation ($\hbar = k_B = 1$):
\begin{eqnarray} \label{eq:master}
    \dot{\rho_t} = \mathcal{L}(\rho_t) = -i [H, \rho_t] + \sum_{k=1}^K \mathcal{D}[L_k] \rho_t,
\end{eqnarray}
where $\rho_t$ is the density operator of the system, $H$ its Hamiltonian, and $\mathcal{D}[L] \rho = L \rho L^\dagger - \{L^\dagger L, \rho\}/2$ a dissipator describing irreversible processes triggered by the environment associated to a set of Lindblad or jump operators $\{L_k\}$ (emission and absorption of quanta, dephasing, etc). Both $H$ and $L_k$ can be time-dependent, following the externally imposed variation of some parameter $\lambda$ that follows a prescribed control protocol $\Lambda := \{ \lambda(t) ; 0 \leq t \leq \tau \}$ up to some final time $\tau$. We assume local detailed balance for the jump operators, i.e. every jump process is related to its inverse counterpart as $L_{\tilde k} = L_k^\dagger e^{-\Delta s_k/2}$, which is also included in the set $\{L_k \}$~\cite{Manzano2018b}. Here $\Delta s_k$ is the entropy change in the environment associated to the $k$th jump (processes represented by hermitian operators are their self-reversed counterparts and hence $\Delta s_k = 0$).

We further assume the system of interest to be initially prepared at time $t=0$ in a pure state $\ket{n}_0$ with probability $p_n(0)$, as sampled from $\rho_0 = \sum_n p_n(0) \ket{n}\bra{n}_0$. Moreover, we introduce a final projective measurement at time $\tau$ on the system using an arbitrary set of rank-1 projectors $\{\ket{m}\bra{m}_\tau\}$, that is useful for the thermodynamic description at the level of fluctuations~\cite{Manzano22}.

The dynamics described by the master equation~(\ref{eq:master}) can be \emph{unravelled} 
into quantum trajectories by introducing a continuous monitoring scheme~\cite{Wiseman2009}, where a generalized measurement is performed on the system at every infinitesimal instant of time $dt$, such that $\rho_{t + dt} = \sum_k M_k \rho_t M_k^\dagger$ with measurement operators verifying $\sum_k M_k^\dagger M_k = \openone$. For concreteness, we assume in the following a quantum jump unravelling, although the results derived here are generically valid for other schemes. Within this approach, the system evolution can be described by a sequence of smooth evolution periods intersected by abrupt \emph{jumps} associated with operators $L_k$, occurring at stochastic times. More precisely, we have 
$M_k = \sqrt{dt} L_k$ with $k=1...K$, for a detection of a jump of type $k$ in the interval $[t , t+ dt]$, and $M_0 = \openone - i H dt - dt \sum_k L_k^\dagger L_k/2$, for no-jumps during $dt$. 

The above procedure is known to describe the state of the system conditioned on a given  record of jumps detected during the evolution as a pure state $\ket{\psi}_t$, following a stochastic Schr\"odinger equation~\cite{Wiseman2009,Carmichael1993} (more details are provided in  Appendix~\ref{app1}). Including the initial state and the final projection, we hence define the complete measurement record up to the final time $\tau$ as $\Gamma_{[0,\tau]} := \{n , (t_1, k_1), (t_2, k_2), ... , (t_J, k_J), m \}$, where $J$ jumps have been detected. Taking the average over different measurement records, we recover the evolution described by the master equation~\eqref{eq:master}.

In an ideal setting, each jump in the system trajectory matches a corresponding detection event. However, in any realistic monitoring setup, many jumps might not be detected. Examples comprise prototypical photon emission from cavities with imperfect mirrors~\cite{Carmichael1993,Wiseman2009,Kimble77}, Ramsey interferometry in maser-like cavity QED~\cite{Benson94,Gleyzes07}, real-time monitoring of tunnelling electrons~\cite{Lu03,Petta04,Bylander05,Sukhorukov07}, or circuit QED setups~\cite{Murch2013,Huard16}. In such situations, the monitoring scheme needs to be modified to take into account the detection efficiency $\eta_k$ of each process $L_k$.

As a consequence of informational leakage in the detection, the state of the system conditioned to a given measurement record can no longer be described by a pure state during the stochastic evolution, being instead a mixture modeled by a density matrix $\sigma_t$. The evolution of the state under imperfect monitoring follows a stochastic master equation of the form:
\begin{eqnarray} \label{eq:smaster}
    d \sigma_t &= -i [H,\sigma_t]dt +\sum_{k} \Big( dt~ (1-\eta_k) \mathcal{D}[L_k] \sigma_t \nonumber \\
               &+ dt~ \eta_k \mathcal{H} [L_k] \sigma_t + dN_{k}~ \mathcal{J}[\sqrt{\eta_k} L_k] \sigma_t \Big),
\end{eqnarray}
where we introduced the superoperators $\mathcal{H}[L] \sigma := \tr[L^\dagger L \sigma] \sigma - \{L^\dagger L, \sigma \}/2$ and $\mathcal{J}[L] \sigma := L \sigma L^\dagger/\tr[L^\dagger L \sigma] -  \sigma$, describing respectively the smooth evolution of the system when no-jumps are detected and the abrupt changes produced by the jumps. Here above the stochastic jumps are incorporated by using Poisson increments $dN_k=\{0, 1\}$ associated to the number of detected jumps $N_k$, which verify $\langle dN_k \rangle = \tr[L_k^\dagger L_k \rho_t] dt$ and $dN_k dN_l = dN_k \delta_{k,l}$~\cite{Wiseman2009}.

For perfect detection efficiency, $\eta_k = 1~ \forall k = 1,..., K$, the second term in the first line of Eq.~\eqref{eq:smaster} disappears, and we recover the ideal monitoring case, for which $\sigma_t = \ket{\psi}\bra{\psi}_t$, and Eq.~\eqref{eq:smaster} is equivalent to the stochastic Schr\"odinger equation as given in App.~\ref{app1}. On the other side, for no-detection $\eta_k = 0~ \forall k$, the two last terms vanish and we recover the Lindblad master equation~\eqref{eq:master}. In this sense, Eq.~\eqref{eq:smaster} interpolates between these two extremes of getting either complete or zero information about the microscopic processes $L_k$.

The stochastic master equation~\eqref{eq:smaster} generates imperfect measurement records $\gamma_{[0,\tau]} := \{n, (t'_1, k'_1), ..., (t'_V, k'_V), m \}$ with $V \leq J$ \emph{visible} jumps, that contain only partial information with respect to the ideal detection case, i.e. $\gamma_{[0,\tau]} \subseteq \Gamma_{[0,\tau]}$. This observation can be made more precise by formally duplicating the original set of $K$ Lindblad operators  $\{L_k\}$ in two sets corresponding, respectively, to visible processes $\{ L_k^\prime := \sqrt{\eta_k} L_k\}$ and their ``hidden" counterparts $\{{L}_k^\ast := \sqrt{1- \eta_k} L_k\}$. In this way the full ideal measurement record can be rewritten as $\Gamma_{[0,\tau]}=\{ n, (t'_1, k'_1)...,(t'_V, k'_V), (t^*_1, k^*_1)...,(t^*_H, k^*_H), m \}$, in terms of both visible $(t'_j, k'_j)$ and hidden $(t^\ast_j, k^\ast_j)$ jumps, with $V + H = J$. That is, $\Gamma_{[0,\tau]} = \gamma_{[0,\tau]} \cup h_{[0,\tau]}$ with $h_{[0,\tau]} := \{ (t^*_1, k^*_1),...,(t^*_H, k^*_H) \}$ the \emph{hidden jumps} that the (imperfect) monitoring scheme fails to detect.

\section{Irreversibility and dissipation} \label{sec:3}
In order to characterize the thermodynamics of monitored processes we employ the total stochastic entropy production, accounting for irreversibility and dissipation at the most general level~\cite{Landi21}. In the case of ideally monitored quantum systems it reads~\cite{Manzano22}:
\begin{eqnarray} \label{eq:Stot}
    S_\mathrm{tot}(\tau) = \ln \left( \frac{\mathbb{P}(\Gamma_{[0,\tau]})}{\tilde{\mathbb{P}}(\tilde{\Gamma}_{[0,\tau]})} \right) = \Delta S_\mathrm{sys}(\tau) + \sum_{r=1}^R \beta_r Q_r(\tau),~~~
\end{eqnarray}
where $\mathbb{P}(\Gamma_{[0,\tau]})$ is the probability to obtain a measurement record $\Gamma_{[0,\tau]}$ in the original process (under the driving protocol $\Lambda$), and $\tilde{\mathbb{P}}(\tilde{\Gamma}_{[0,\tau]})$ the probability to obtain the time-reversed sequence of (inverse) jumps $\tilde{\Gamma}_{[0,\tau]}$, when the external driving protocol is also time-reversed, i.e. under $\tilde{\Lambda} = \{ \lambda(\tau -t) ; 0 \leq t \leq \tau \}$ (for more details see App.~\ref{app1}). In the second equality we split the entropy production in the stochastic change in system entropy (self-information) $\Delta S_\mathrm{sys}(\tau) = -\ln p_m(\tau) + \ln p_n(0)$ and the total accumulated heat dissipated during the trajectory, with $Q_r(\tau) = \int_0^\tau \sum_{k \in \mathcal{K}_r} dN_k \Delta s_k T_r$ the heat transferred into reservoir $r$ at inverse temperature $\beta_r= 1/ T_r$, through the associated set of processes $\mathcal{K}_r$. Notice that while the total stochastic entropy production above depends on the whole trajectory $\Gamma_{[0,\tau]}$ we used for convenience the short-hand notation $S_\mathrm{tot}(\tau)$. 

Equation~\eqref{eq:Stot} follows from the micro-reversibility of the monitored dynamics at the level of single trajectories~\cite{Manzano2018b,Manzano22}. On the other hand, when imperfect detection is considered, information leakages will generally hinder micro-reversibility. Nevertheless, we can define an effective information-theoretical indicator of irreversibility as:
\begin{eqnarray} \label{eq:Sigma}
    \Sigma(\tau) := \ln \left( \frac{\mathrm{P}(\gamma_{[0,\tau]})}{\tilde{\mathrm{P}}(\tilde{\gamma}_{[0,\tau]})} \right) = \Delta S_\mathrm{sys}(\tau) + \phi(\tau), 
\end{eqnarray}
with marginalized path probabilities $\mathrm{P}(\gamma_{[0,\tau]}) := \sum_{h_{[0,\tau]}}\mathbb{P}(\Gamma_{[0,\tau]})$ and $\tilde{\mathrm{P}}(\tilde{\gamma}_{[0,\tau]}) := \sum_{h_{[0,\tau]}} \tilde{\mathbb{P}}(\tilde{\Gamma}_{[0,\tau]})$, obtained by summing over all possible hidden jumps sequences  $h_{[0,\tau]}$.
Using the fact that probabilities of initial states in both forward and time-reversed processes do not depend on the monitoring efficiencies ${\eta_k}$, the second equality in \eqref{eq:Sigma} follows from the introduction of {the effective entropy flux $\phi(\tau) := \ln [\mathrm{P}(\gamma_{[0,\tau]}| n)/\tilde{\mathrm{P}}(\tilde{\gamma}_{[0,\tau]}|m)]$}.  However it is important to notice that there is not any a priori relation of $\phi(\tau)$ with the actual thermodynamic entropy exchanges with the environment, i.e., with the heat exchanged with each reservoir divided by its temperature. {Details regarding the construction of the information-theoretical irreversibility estimator and their properties are given in Appendix~\ref{app5}.}

The form of $\Sigma(\tau)$ in Eq.~\eqref{eq:Sigma} immediately implies its non-negativity on average as $\langle \Sigma(\tau) \rangle = D[\mathrm{P}(\gamma_{[0,\tau]}) || \mathrm{P}(\tilde{\gamma}_{[0,\tau]})] \geq 0$ is a Kullback-Leibler divergence between path probabilities~\cite{Cover2006}, in analogy to the classical case~\cite{Kawai2007}. 
Moreover, an integral fluctuation theorem is verified by $\Sigma$ itself as: 
\begin{equation}
\langle e^{-\Sigma} \rangle = \sum_{\gamma_{[0,\tau]}} \tilde{\mathrm{P}}(\tilde{\gamma}_{[0,\tau]}) = 1,
\end{equation}
{where the average is over visible trajectories $\gamma_{[0,\tau]}$ [in contrast with Eq.~\eqref{eq:FT} where the average is only over hidden jumps]}. However, the relation between the information-theoretical quantity $\Sigma$ with actual thermodynamic quantities along visible trajectories, and in particular with the underlying stochastic entropy production $S_\mathrm{tot}$ in Eq.~\eqref{eq:Stot} was, so far, unknown.

Our main results presented above, Eqs.~\eqref{eq:FT} and \eqref{eq:bound}, provide a clean link between the apparent irreversibility in the imperfectly monitored process as measured by $\Sigma$ and the underlying (thermodynamic) dissipation as measured by the total entropy production $S_\mathrm{tot}$ (detailed proofs are provided in Appendix~\ref{app2}). The key ingredient to obtain the coarse-graining fluctuation theorem in Eq.~\eqref{eq:FT} is the use of conditional path probabilities for the hidden jumps with respect to visible ones, namely $\mathrm{P}(h_{[0,\tau]} | \gamma_{[0,\tau]}) := \mathbb{P}(\Gamma_{[0,\tau]})/\mathrm{P}(\gamma_{[0,\tau]})$, which allows us to write conditional averages of arbitrary functionals $X(\Gamma_{[0,\tau]})$ of the full measurement records as $\langle X | \gamma_{[0,\tau]} \rangle = \sum_{h_{[0,\tau]}} \mathrm{P}(h_{[0,\tau]} | \gamma_{[0,\tau]}) X(\Gamma_{[0,\tau]})$. 

The fluctuation theorem~\eqref{eq:FT} can be further rewritten in standard form as $\langle e^{-(S_\mathrm{tot}- \Sigma)}| \gamma_{[0,\tau]}\rangle = 1$, from which we obtain, inspired by Refs.~\cite{Jarzynski2011,Jarzynski08}, the following bound relating the fluctuations of $S_\mathrm{tot}$ with $\Sigma$:
\begin{equation} \label{eq:min}
 \mathrm{Pr}(S_\mathrm{tot} - \Sigma < - \xi) \leq e^{-\xi},
\end{equation}
with $\xi \geq 0$. The above inequality implies that the probability of having stochastic entropy production values $S_\mathrm{tot}$ below the estimation $\Sigma$ given the visible measurement record $\gamma_{[0,\tau]}$, is exponentially suppressed. In other words, the probability that $\Sigma$ overestimates stochastic entropy production in the monitored thermodynamic process is exponentially negligible.

\begin{figure*}[t]
    \centering
    \includegraphics[width=1.0\linewidth]{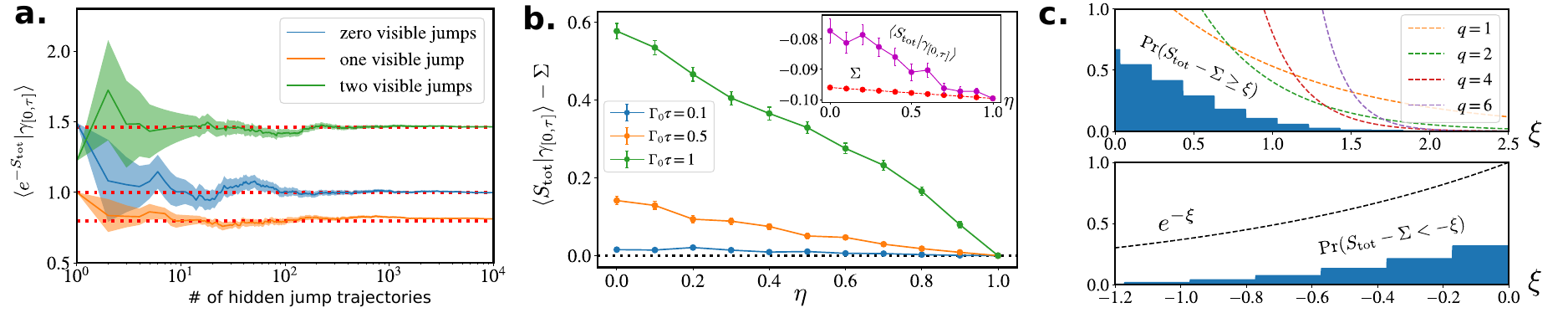}
    \caption{Numerical testing of the main results. a) Convergence of fluctuation theorem in Eq.~\eqref{eq:FT} by conditional sampling of $\langle e^{-S_\mathrm{tot}} | \gamma_{[0,\tau]}\rangle$ over hidden trajectories (solid lines) to $e^{-\Sigma}$ (dashed lines) for three different cases of visible trajectories with zero, one, or two jumps (see legend). Statistical errors are given by shaded areas around the lines. b) Test of inequalities~\eqref{eq:bound} and \eqref{eq:heat} as a function of the detection efficiency in the symmetric case, $\eta := \eta_+=\eta_-$, for a trajectory without visible jumps with different lengths (see legend).
    The main plot shows the difference between the conditional entropy production $\langle S_\mathrm{tot} | \gamma_{[0,\tau]\rangle}$ and the information-theoretical irreversibility indicator $\Sigma$, while the two quantities are separately represented in the inset. c) Upper (top) and lower (bottom) statistical bounds for the estimation of entropy production through $\Sigma$ as a function of the deviations $\xi$ from Eqs.~\eqref{eq:boundsq} and \eqref{eq:min} respectively, using different values of the exponent $q\geq 1$. Other parameters: $\Gamma_0 = 10^{-3} \omega$, $\epsilon = 10^{-2} \omega$, $\beta \omega = 1/5$ and $\hbar \omega = 1$ (for all plots). For (a) $\Gamma_0 \tau = 1$, $\eta_{+} = \eta_{-} = 0.2$. {For (b) inset $\Gamma_0 \tau=0.1$.} For (c) $\Gamma_0 \tau = 3$, $\eta_{-} = 0.5$, $\eta_{+} = 0.2$.}
    \vspace{-0.3cm}
    \label{fig:results}
\end{figure*}

Similarly, we provide a family of bounds for the right tail of $S_\mathrm{tot}$ distribution:
\begin{eqnarray} \label{eq:boundsq}
\mathrm{Pr}(S_\mathrm{tot} - \Sigma \geq  \xi) \leq  e^{- q \xi} \langle e^{q (S_\mathrm{tot}-\Sigma)} | \gamma_{[0,\tau]} \rangle,    
\end{eqnarray}
with {$q \geq 1$ and} again $\xi \geq 0$. In the limit of large trajectories, $\tau \rightarrow \infty$, the right hand side of the above inequality can be minimized~\cite{Manzano22b} for $q=1$, leading to 
 $\mathrm{Pr}(S_\mathrm{tot} \geq \Sigma + \xi) \leq  e^{-\xi} e^{\tau K(1)}$, with $K(q):=\lim_{\tau \rightarrow \infty} \ln \langle e^{q (S_\mathrm{tot}(\tau) - \Sigma)} | \gamma_{[0, \tau]} \rangle / \tau$ a scaled cumulant generation function~\cite{Touchette09}.
The above Eq.~\eqref{eq:boundsq} and its long time limit thus provide us precise bounds on how much $S_\mathrm{tot}$ may be underestimated by $\Sigma$, following an exponential decay on $\xi$, attenuated by the factor $\langle e^{q (S_\mathrm{tot}-\Sigma)} | \gamma_{[0,\tau]} \rangle \geq 1$. As a corollary we observe that only if $K(1) \rightarrow 0$ the inequality $\eqref{eq:bound}$ becomes tight for arbitrary large trajectories. For detailed proofs of Eqs.~\eqref{eq:min} and~\eqref{eq:boundsq} and the long-time limit, see Appendix~\ref{app3}.

The explicit expressions of stochastic entropy production in Eq.~\eqref{eq:Stot} and the irreversibility indicator in Eq.~\eqref{eq:Sigma}, allow us to rewrite the bound in Eq.~\eqref{eq:bound} in terms of the heat as:
\begin{eqnarray} \label{eq:heat}
    \phi(\tau) \leq  \sum_r \beta_r \langle Q_r(\tau) | \gamma_{[0,\tau]} \rangle,
\end{eqnarray}
where $\langle Q_r | \gamma_{[0,\tau]} \rangle = \sum_{h_{[0,\tau]}} \mathrm{P}(h_{[0,\tau]} | \gamma_{[0,\tau]}) Q_r(\tau)$ is the expected heat dissipated into reservoir $r$ given the visible measurement record $\gamma_{[0,\tau]}$. We notice that here $\phi(\tau)$, as well as $\langle Q_r | \gamma_{[0,\tau]} \rangle$ can be either positive or negative (a explicit expression for $\langle Q_r(\tau) | \gamma_{[0,\tau]} \rangle$ for quantum-jump trajectories is given below). However $\phi(\tau)$ always provides us a lower bound on heat dissipation and hence can be regarded as the minimum (integrated) entropy flow to the environment compatible with observations $\gamma_{[0,\tau]}$.  
The equality case in Eq.~\eqref{eq:heat}, as well as in Eq.~\eqref{eq:bound}, is reached for unit efficiencies $\eta_k = 1~~\forall k$, in which case $\gamma_{[0,\tau]} = \Gamma_{[0,\tau]}$. However, Eq.~\eqref{eq:heat} provides a tight bound on the real heat whenever $Q_r(\tau)$ remains similar for every possible set of hidden jumps $h_{[0,\tau]}$, i.e. when the contribution to the total heat from the hidden jumps is small. That would be typically the case e.g. for short times, small energies of the inefficiently detected channels, or high temperatures. 

It is also worth noticing that by taking the average in Eq.~\eqref{eq:heat} or Eq.~\eqref{eq:FT} over final and (or) initial measurement outcomes $\{m, n\}$, other equalities and inequalities can be obtained that depend only on the detected jumps sequence $\{(t_1^\prime, k_1^\prime),..., (t^\prime_V, k^\prime_V)\}$. {In Appendix~\ref{app6} we provide extra fluctuation relations and second-law inequalities following that procedure.} On the other hand, by taking the average over the entire visible records $\gamma_{[0,\tau]}$ we obtain, as expected, $\langle \phi(\tau) \rangle \leq \sum_r \beta_r \langle Q_r(\tau) \rangle$.

\subsection*{Conditional heat for quantum-jump trajectories } \label{sec:4}
For the case of quantum jump trajectories the explicit expression for the conditional expectation of the heat dissipated into reservoir $r$ can be written down in terms of the stochastic increments $dN_k$ appearing in the stochastic master equation~\eqref{eq:master} as:
\begin{align} \label{eqs:condheat}
  &\langle Q_r(\tau) | \gamma_{[0,\tau]} \rangle = \mathrm{P}[m|v_{[0,\tau]},n]~T_r \\
  &\int_0^\tau  \sum_{k \in \mathcal{K}_r} \Big( dN_k \Delta s_k + dt (1 - \eta_k) \tr[L_k^\dagger L_k \sigma_t] \Delta s_k \Big), \nonumber 
\end{align}
where the multiplicative factor $\mathrm{P}[m|v_{[0,\tau]},n]$ represents the conditional probability to obtain outcome $m$ in the final measurement given initial state $n$ and a recorded sequence of visible jumps $v_{[0,\tau]} = \{(t^\prime_1, k^\prime_1),...,  (t^\prime_V, k^\prime_V)\}$. The first term inside the integral above arises from heat exchanges as a consequence of detected (visible) jumps, and the second is the expected heat dissipation from hidden jumps, which is generically non-zero (for an illustration see Fig.~\ref{fig:sketch}b). Notice that by summing $m$ in both sides of Eq.~\eqref{eqs:condheat}, the factor in the r.h.s. disappears, $\sum_m \mathrm{P}[m|v_{[0,\tau]},n] = 1$, and we obtain in the l.h.s. the conditional heat (conditioned on the initial state and the detected jumps) independently of the final state $m$ of the trajectory, namely $\langle Q_r(\tau) | v_{[0,\tau]},n \rangle$.

For the case of energy (but no particle) exchanges with the environment, the Lindblad operators verify $[H_S, L_k] =  \Delta E_k L_k$, with $H_S$ the (bare) system Hamiltonian and $\Delta E_k$ the energy transferred to the environment in the jump~\cite{Manzano22}. Then we have $\Delta s_k = \Delta E_k/T_r$ and in the second term above $\tr[L_k^\dagger L_k \sigma_t] \Delta s_k = \tr[H_S \mathcal{D}[L_k]\sigma_t]$. In that case we get:
\begin{align}\label{eq:condheat2}
 &\langle Q_r(\tau) | \gamma_{[0,\tau]} \rangle = \mathrm{P}[m|v_{[0,\tau]},n] \\ &\int_0^\tau  \sum_{k \in \mathcal{K}_r} \Big( dN_k \Delta E_k + dt (1 - \eta_k) \tr[H_S \mathcal{D}[L_k]\sigma_t] \Big), \nonumber
\end{align}
where again summing over $m$ in both sides leads to the disappearance of the conditional probability $\mathrm{P}[m|v_{[0,\tau]},n]$. We remark that for $\eta_k = 1~\forall k$ only the first term survives and we recover the  expression for ideal quantum jump trajectories~\cite{Manzano22}. On the other hand, if $\eta_k = 0~ \forall k$, the first term vanishes since the stochastic variables $dN_k = 0$ at every time, and we recover from the second term the known expressions for average heat in weakly coupled open quantum systems~\cite{Spohn1978,Alicki1979}.

\section{Illustrative example} \label{sec:5}
To exemplify our results, we have chosen a simple two-level system interacting with a bosonic environment at thermal equilibrium and weakly driven by a resonant coherent field. The Hamiltonian for the two-level system is $H_0 = \omega \ket{1}\bra{1}$, which exchanges energy quanta with the environment through emission and absorption processes. These are described by the Lindblad operators $L_- = \sqrt{\Gamma_0(\bar{n}+1})\sigma_-$ and $L_+ = \sqrt{\Gamma_0\bar{n}}\sigma_+$, where $\sigma_-$ and $\sigma_+$ are lowering and raising operators, $\Gamma_0$ is the spontaneous decay rate, and $\bar{n}$ is the average number of bosons in the reservoir. The driving on the system is described by a time-dependent perturbation to $H_0$, namely $V(t) = \epsilon (\sigma_+ e^{-i\omega t} +\sigma_- e^{i\omega t})$, with $\epsilon \ll \omega$, which ensures that the structure of the Lindblad operators is unaffected by the driving~\cite{Manzano2018b, Lopez2023, Trushechkin2016}. { Notice that we also choose the driving to be resonant with the system's energy spacing, which is the easiest way for the system to take energy from the driving.}

{
By moving to a rotating frame at angular frequency $\omega$, we might be able to remove the time dependency in the total  Hamiltonian $H(t) = H_0 + V(t)$, so that the driving term  in the interaction picture with respect to $H_0$ is $\epsilon \sigma_x$, with $\sigma_x$ the Pauli matrix. In this picture the evolution of the system can be described by the Lindblad master equation~\eqref{eq:master} with Hamiltonian term $\epsilon \sigma_x$ and the Lindblad operators $L=\{L_{-}, L_{+}\}$ introduced above. They verify $[H_0, L_{\pm}] = \pm \omega L_{\pm}$ and the local detailed balance relation $L_- = e^{- \omega/(2T)} L_+^{\dagger}$, leading to stochastic entropy exchanges with the environment $\Delta s_{\pm} = \pm \omega/T$. 
}

\begin{figure*}[tbh]
    \centering
    \includegraphics[width=1.0\linewidth]{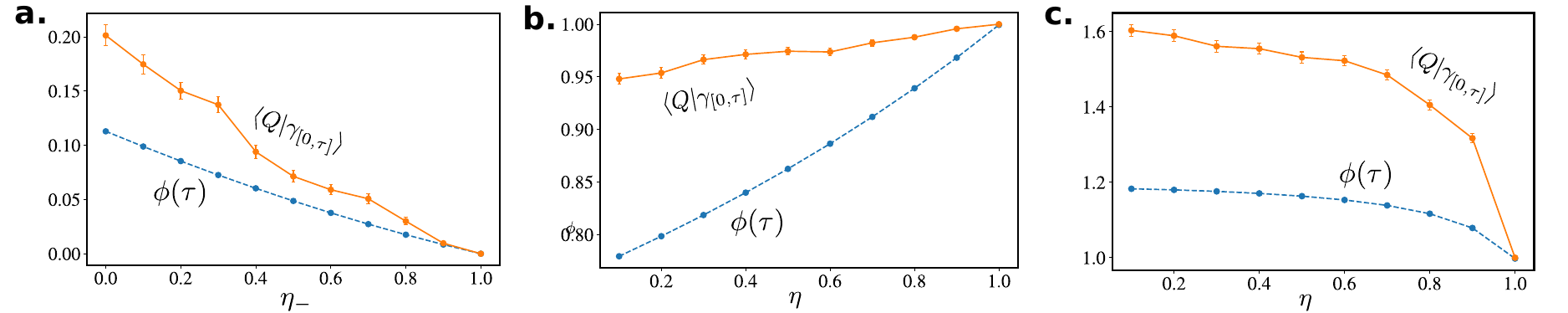}
    \caption{ Numerical testing of inequality~\eqref{eq:heat} as a function of detection efficiency for different trajectories. a) Trajectory with no visible jumps in an interval of duration $\Gamma_0 \tau = 0.5$ and fixed absorption efficiency $\eta_+ = 1$, b) trajectory of total length $\Gamma_0 \tau = 0.1$ with one visible emission at $\Gamma_0 t_1=0.05$, and equal detection efficiencies $\eta := \eta_+ = \eta_{-}$ c) trajectory of total length $\Gamma_0 \tau = 0.5$ with one visible emission at $\Gamma_0 t_1=0.25$ and equal detection efficiencies. Parameters: $\Gamma_0 = 10^{-3} \omega$, $\epsilon = 10^{-2} \omega$, $\beta \omega = 1/5$ and $\hbar \omega = 1$.}
    \vspace{-0.3cm}
    \label{fig:supplot}
\end{figure*}

We focus on steady-state conditions, where the system reaches a nonequilibrium state showing coherence in the energy basis. This state is maintained by the continuous dissipation of input work from the drive as heat into the thermal environment, which is partially monitored (see Fig.~\ref{fig:sketch}b). We numerically tested our main fluctuation relations and inequalities by assuming imperfect detection of both emission and absorption processes, with respective efficiencies $\eta_{-}$ and $\eta_{+}$. Quantum trajectories of this system conditioned on visible jumps can be computed from the stochastic master equation Eq.~\eqref{eq:smaster}, while evaluation of conditional averages requires modifications on the original quantum-trajectory Monte Carlo algorithm (see App.~\ref{app7}). We find an excellent convergence for the fluctuation theorem in Eq.~\eqref{eq:FT} when increasing the number of sampled hidden jump sequences $h_{[0,\tau]}$ for three different fixed (visible) records $\gamma_{[0,\tau]}$(Fig.~\ref{fig:results}a). {There relatively short trajectories have been used so that the convergence of the FT can be better visualized without an excessive number of hidden jumps samples.} In Fig.~\ref{fig:results}b we show the tightness of the bounds~\eqref{eq:bound} and \eqref{eq:heat} on the heat dissipated into the environment when modifying the efficiency of detection. As we can see there, the difference $\langle S_\mathrm{tot} | \gamma_{[0,\tau] \rangle} - \Sigma = \beta \langle Q | \gamma_{[0,\tau]} \rangle - \phi$ approaches zero when increasing the efficiency of detection of both emission and absorption events, leading to the saturation of the bound for perfect detection. Moreover, as shown from the different lines, when increasing the length of the trajectories for a fixed visible jump record, this difference becomes greater due to the highest occurrence of hidden jumps. We further evaluated the statistics of $S_\mathrm{tot}$ over hidden jump sequences (Fig.~\ref{fig:results}c) in order to test the exponential decay bound of entropy production fluctuations below $\Sigma$ predicted by Eq.~\eqref{eq:min} (bottom plot) and the attenuated decay above $\Sigma$ bounded by the family of curves in Eq.~\eqref{eq:boundsq} (top plot).

In Fig.~\ref{fig:supplot} we show examples of the environmental contribution of the estimator $\Sigma(\tau)$, namely $\phi(\tau)$ in Eq.~\eqref{eq:heat}, as compared with the conditional average of dissipated heat $\langle Q (\tau) | \gamma_{[0,\tau]}\rangle$ in Eq.~\eqref{eqs:condheat} for different visible trajectories, complementing the previous plots. Fig.~\ref{fig:supplot}a corresponds to a trajectory with no visible jumps during an interval $\Gamma_0 \tau = 0.5$ for fixed $\eta_+=1$, while in Figs.~\ref{fig:supplot}b and \ref{fig:supplot}c we consider trajectories with one visible jump (down) and lengths $\Gamma_0 \tau= 0.1$ and $\Gamma_0 \tau=0.5$ respectively, for symmetric detection efficiencies $\eta_+ = \eta_- = \eta$. While in the three cases we actually observe different shapes for the curves, the estimator $\phi(\tau)$ always follows $\langle Q (\tau) | \gamma_{[0,\tau]}\rangle$, both of them becoming closer for higher efficiencies. However, the maximum distance between them may vary in general depending on the specific (stochastic) trajectory under consideration.

\section{Experimental implementation and Thermodynamic Inference} \label{sec:6}
The irreversibility estimator $\Sigma(\tau)$ can be calculated with the help of the stochastic master equation~\eqref{eq:smaster} and its time-reversed counterpart. In that case, knowledge of the system Hamiltonian, Lindblad operators, and the efficiency of dissipative channels is required, as it is generically the case in many experiments~\cite{Murch2013,Huard2016,Naghiloo2018,Rossi2019,Minev2019}, where it could be directly compared to the conditional entropy production $\langle S_\mathrm{tot} | \gamma_{[0,\tau]} \rangle = \Delta S_\mathrm{sys}(t) + \sum_r \beta_r \langle Q_t(\tau) | \gamma_{[0,\tau]} \rangle$ using the expressions for the conditional heat in Eqs.~\eqref{eqs:condheat} and \eqref{eq:condheat2}. 
The explicit expression for the estimator $\Sigma(\tau)$ is obtained in Appendix~\ref{app5} from the (marginal) path probabilities for visible trajectories in the forward and time-reversed processes.

On the other hand, $\Sigma(\tau)$ can also be used as a tool for thermodynamic inference purposes in situations where the details of the dynamics, or even the particular efficiency of the channels are not known. In that case, $\Sigma(\tau)$ might be directly evaluated from the statistics of experimental samples of the monitored system by implementing both the forward and time-reversed processes. In this context, it is important to stress that our results predict that in the visible time-reversed process there is an effective exchange of efficiencies between reciprocal channels related by local detailed balance (see App.~\ref{app5} for details). That is, implementing the time-reversed process under imperfect detection not only requires reversing the driving protocol $\Lambda \rightarrow \tilde \Lambda$ as in the ideal case, but also to modify the detection efficiency of each channel to match the reciprocal one, $\eta_k \rightarrow \eta_{\tilde{k}}$. This is a key feature of imperfect monitoring schemes following from the microreversibility principle as applied to the ideal (extended) scenario. It naturally prevents the appearance of divergences in the irreversibility estimator ---and consequently in the entropy production--- for the cases in which only one of a pair of reciprocal channels can be detected with non-zero efficiency (e.g. if only emitted photons can be detected but not absorptions). This property has been unnoticed so far and applies in both transient and stationary regimes.

The extra property of the exchange of efficiencies between reciprocal channels in the time-reversed process may at the same time pose some challenges for the practical implementation of the time-reversed process in the laboratory in some setups, e.g., for direct photodetection of emitted quanta with no possibility of detecting absorptions. Indeed, the experimental setup for detecting the quantum jumps may require some flexibility to allow the implementation of both the forward and time-reversed processes, so that the irreversibility estimator $\Sigma$ can be directly obtained from the likelihood of the sampled trajectories in both processes. We notice that in general, there is no problem in implementing an efficiency that is lower than the one achieved by the real detector, since this can be done by just neglecting some of the detected jumps so that the effective efficiency matches the desired value. However engineering a higher efficiency could be more challenging and require more sophisticated detection techniques beyond simple photodetection of the transition of interest. 

A relevant example where more sophisticated detection techniques are employed that would allow the direct use of $\Sigma$ for thermodynamic inferece are setups where the monitoring scheme uses an extra auxiliary level connected to the ones in which the jumps are to be detected. These schemes have been employed since the first experimental observations of quantum jumps in atomic physics in the 80s under the name of shelving scheme~\cite{Dehmelt86,Toschek86,Wineland86,Basch95}, where intermittencies in the fluorescence measurement signal of a third auxiliary level transition allowed to detect quantum jumps between two levels of interest. Similar techniques have been employed more recently for detecting spin-dependent quantum jumps in quantum dots~\cite{Vamivakas10} or to catch and reverse quantum jumps in superconducting atoms~\cite{Minev2019}.

Particularly, according to previous proposals~\cite{Carvalho11,Elouard2017b} where the dissipative environment was engineered, in the prototypical case of a monitored two-level system, the addition of an auxiliary metastable level $\ket{m}$ with higher energy allows the detection of jumps between the ground $\ket{g}$ and exited $\ket{e}$ levels by only detecting emitted photons of two different frequencies. In those proposals, the extra level is connected to the ground state through a nearly resonant (weak) laser driving, which allows the adiabatic elimination of the extra state due to its short lifetime, but permits at the same time the direct detection of photoemissions from the metastable to the excited state, $\ket{m} \rightarrow \ket{e}$, leading to an effective jump $\ket{g} \rightarrow \ket{e}$, using an extra photodetector. {Realistic values for the efficiency in direct detection of fluorescence using e.g. an avalanche photodiode are around $\eta = 0.8$~\cite{Warszawski02}}. In such a setup, the inversion of the efficiencies needed to implement the time-reversed process discussed in this paper can be conveniently achieved by simply exchanging the photodetector used for detecting the emissions in the transition $\ket{m} \rightarrow \ket{e}$ with the one used for detecting the emissions in the transition $\ket{e} \rightarrow \ket{g}$. {That implementation would allow the experimental sampling of trajectories in both forward and time-reversed processes. The irreversibility estimator $\Sigma$ could then be constructed from these statistics by evaluating the relative frequency of observation of the visible trajectories [see Eq.~\eqref{eq:Sigma}] , without the need of any detail of the dynamical evolution (not even the efficiency of  the channels). Using inequality~\eqref{eq:bound} this estimator thus provides a reliable lower bound on the dissipation incurred in generic thermodynamic processes operated over the two-level system under investigation at the level of single trajectories.}

\section{Discussion and conclusions} \label{sec:7}
We obtained universal fluctuation relations, as presented in Eqs.~\eqref{eq:FT}, \eqref{eq:min} and \eqref{eq:boundsq}, for the entropy production generated in imperfectly monitored quantum systems, that are conditioned on the registered measurement record. These relations apply for generic initial states and Markovian (quantum) processes arbitrarily far from equilibrium.
We have shown that despite a part of the information on the energy and matter exchanges with the environment may be hidden to the observer, a lower bound on the entropy production in the process can be obtained [Eq.~\eqref{eq:bound}], via a suitable indicator of irreversibility computable along observed trajectories, $\Sigma$ in Eq.~\eqref{eq:Sigma}. Remarkably, this indicator can be computed at the level of single (imperfectly observed) trajectories, allowing us to not only estimate the average dissipation of a given process and obtain bounds on all the even moments of its distribution, but also to asses the minimum dissipation associated to particularly interesting events, such as e.g. rare events.

Importantly, our results apply to classical and quantum monitored systems alike. In the classical limit, where the entire evolution occurs within the system energy basis associated to the Hamiltonian $H$, quantum jumps reduce to classical jumps between (eventually degenerate) energy levels, as described by stochastic Markovian dynamics in continuous time, where some of those jumps might not be detected by the observer. Within the classical context, our results extend previous techniques for the  estimation of average entropy production in non-equilibrium steady-states~\cite{Roldan10,Martinez19} to the stochastic entropy production along single visible trajectories in generic Markovian dynamics. This opens the door for the estimation and test of dissipation at the level of fluctuations in energy-transduction processes in living and non-living active matter at the microscale~\cite{Skinner21,Lynn21,Roldan21,Seara21,Bisker22}, where information about trajectories is often readily available~\cite{Noji97,Weigel11,Fakhri14}. In this context, it would be also interesting to compare our approach with recent developments for entropy production estimation in situations where only a few transitions between system states can be observed~\cite{Roldan22,Seifert22} or for snippets between Markovian events~\cite{Meer23,Degunther24,Degunther24b}.

On the quantum side, while we focused on quantum jump trajectories, the results presented here are general and may be explicitly extended to the case of diffusive dynamics as well (which can be indeed derived from the quantum jumps approach in a particular limit~\cite{Wiseman2009,Carmichael1993,Najmeh2020}). This is an interesting perspective that we leave for future work, which will allow direct comparison with alternative proposals to define and measure entropy production in monitored quantum systems using phase-space methods~\cite{Belenchia2020,Kewming22}, or with proposals for heuristic arrow-of-time indicators~\cite{Jayaseelan21}. 

We expect that our results will be of relevance in experimental situations aiming to test quantum thermodynamics within the quantum trajectory approach~\cite{Pekola15,Naghiloo2020,Rossi2020,Karimi20a} where the thermodynamic role of imperfect jump detection has not been yet fully explored. Given the growing importance of the energetic footprints in information and communication technologies, it would be also interesting to explore applications of our results for assessing dissipation in quantum technologies of the noisy intermediate-scale quantum (NISQ) era~\cite{Auffeves22}, {such as the determination of thermodynamic computational costs~\cite{Wolpert24} in quantum computation.}

\acknowledgments
We thank Javier Aguilar for interesting discussions regarding conditional sampling of trajectories. We wish to acknowledge support from the María de Maeztu project (CEX2021-001164-M) for Units of Excellence, QUARESC project (PID2019-109094GB-C21), CoQuSy project (PID2022-140506NB-C21) and QuTTNAQMa project (PID2020-117347GB-I00), funded by the Spanish State Research Agency MCIN/AEI/10.13039/501100011033 and FEDER, UE. GM is supported through the Ram\'on y Cajal program (RYC2021-031121-I) funded by MCIN/AEI/10.13039/501100011033 and European Union NextGenerationEU/PRTR. RL acknowledges the financial support by the Grant No. PDR2020/12 sponsored by Comunitat Autonoma de les Illes Balears through the `Direcció General de Política Universitaria i Recerca' with funds from the Tourist Stay Tax Law ITS 2017-006 and the Grant No. LINKB20072 from the CSIC i-link program 2021. MFC also acknowledges funding from Generalitat Valenciana (CIACIF/2021/434).

\appendix

\section{Entropy production in quantum-jumps trajectories} \label{app1}
The unraveling of the master equation in Eq.~\eqref{eq:master} by using an ideal direct detection scheme~\cite{Wiseman2009}, leads to the following stochastic Schr\"odinger equation:
\begin{equation}
\begin{split}
    d\ket{\psi}_{t} = [-i dt H - \frac{dt}{2}\sum_{k} \left( L_k^\dagger L_k - \langle L_k^\dagger L_k \rangle_\psi \right)  
    \\ + \sum_k dN_k \left(\frac{L_k}{\sqrt{\langle L_k^\dagger L_k \rangle}_\psi} - \mathds{1} \right) ] \ket{\psi}_t
\label{eqs:SMEcomplete}
\end{split}
\end{equation}
where we denoted $\langle L_k^\dagger L_k \rangle_\psi = \bra{\psi}_t L_k^\dagger L_k \ket{\psi}_t$. Here the first two terms proportional to $dt$ generated a smooth evolution associated to the periods when jumps are not detected. On the other hand the last term is proportional to the stochastic Poisson increments $dN_k$. When $dN_k = 1$ this last term produces a sharp change in the stochastic wave function $\ket{\psi}_t$ stemming from the occurrence of a quantum jump. The original Lindblad master equation~(4) is recovered from Eq.~\eqref{eqs:SMEcomplete} by constructing the evolution of $\ket{\psi}\bra{\psi}_t$ and taking the average over possible outcomes of the jumps $\langle dN_k \rangle = \tr[L_k^\dagger L_k \rho_t] dt$.

Given an ideal record of $J$ jumps $j_{[0,\tau]} =\{(t_1, k_1)...,(t_J, k_J)\}$ during an interval $[0,\tau]$, which in this case we assume to be perfectly detected, the associated evolution of the system from an arbitrary initial state $\ket{\psi}_0$ up to time $\tau$ can be written as:
\begin{eqnarray} \label{eqs:psi}
    \ket{\psi}_t = \frac{\mathcal{T}(j_{[0,\tau]})}{\sqrt{p_\Gamma}} \ket{\psi}_0,
\end{eqnarray}
where we used the following trajectory evolution operator for the monitored system containing the information for the specific sequence of jump and no-jump periods during the evolution: 
\begin{equation}
 \mathcal{T}(j_{[0,\tau]}) = \mathcal{U}(\tau,t_J)L_{k_J}~...~\mathcal{U}(t_2,t_1)L_{k_1}\mathcal{U}(t_1,0),
\end{equation}
and the normalization $p_\mathcal{T} = \tr [\mathcal{T} \ket{\psi}\bra{\psi}_0 \mathcal{T}^{\dagger}]$ representing the probability of the given jump sequence $j_{[0,\tau]}$ when starting from the initial state $\ket{\psi}_0$. Here above { $\mathcal{U}(t_2,t_1)=T_+ {\rm exp}[-i\int_{t_1}^{t_2}dt(H - i\sum L_k^{\dagger}L_k/2)]$} is the (non-unitary) evolution operator associated to a period with no jumps between $t_1$ to $t_2$.

As mentioned in Sec.~\ref{sec:2}, in order to address the main thermodynamic quantities along trajectories, we also introduce a two-point measurement (TPM) scheme in combination with the continuous monitoring scheme~\cite{Manzano22}. This approach consists of performing projective measurements of arbitrary observables both at the beginning and at the conclusion of the indirectly monitored process. In particular let's assume the system is prepared in a pure state $\ket{n}$ sampling from the initial density operator of the system, $\rho_0 = \sum_n p_n(0) \ket{n}\bra{n}_0$. This is equivalent to performing an initial projective measurement using a complete set of projectors $\{\Pi_n^0 = \ket{n}\bra{n}_0\}$. After that, the evolution of the system proceeds under continuous monitoring until a final time $\tau$, where a second projective measurement is performed in an arbitrary basis using a complete set of projectors $\{\Pi_m^\tau = \ket{m}\bra{m}_\tau \}$ (for simplicity we may consider the basis of $\rho_\tau = \sum_m p_m(\tau) \ket{m}\bra{m}_\tau$). Denoting the outcome of the initial measurement as $n$ and the one of the final measurement as $m$, the complete trajectory of the system can be written as $\Gamma_{[0,\tau]} = \{ n,  (t_1, k_1)...,(t_J, k_J) , m \}$ and its probability to occur (path probability) reads: 
\begin{equation}
    \mathbb{P}(\Gamma_{[0,\tau]}) = p_{n}(0) \tr \left[ \Pi_{m}^\tau \mathcal{T}(j_{[0,\tau]}) \Pi_{n}^0 \mathcal{T}^{\dagger}(j_{[0,\tau]})\right],
    \label{eqs:completeprob}
\end{equation}
where the first term is the probability of sampling $n$ at the beginning, and the second term is the conditional probability to obtain the jump sequence $j_{[0,\tau]}$ and final outcome $m$ in the second projective measurement when starting from $\ket{n}_0$.

The definitions of reversibility and its link with entropy production comprise the comparison of the occurrence of trajectories $\Gamma_{[0,\tau]}$ with the occurrence of their time-reversed counterpart $\tilde{\Gamma}_{[0,\tau]}$ when the driving protocol $\Lambda$ is also inverted. Here, it is convenient to introduce the (anti-unitary) time-reversal operator in quantum mechanics, $\Theta$, which basically changes the sign of the odd variables under time-reversal symmetry in the operators to which it is applied. Moreover, in the following to denominate operators in the time-reversed dynamics, we will use a tilde.

The time-reversed or backward process can be defined from the forward one as follows. It starts with the (inverted) final state of the forward process at time $\tau$, namely $\tilde{\rho}_0 := \Theta \rho_\tau \Theta^\dagger = \sum_m p_m(\tau) \Theta \Pi_m^\tau \Theta^\dagger$ over which a measurement with projectors $\{\Theta \Pi_m^\tau \Theta^\dagger\}$ where outcome $m$ is obtained, then the monitoring procedure is run subjected to the time-reversed driving protocol $\tilde{\Lambda} = \{ \lambda(\tau -t) ; 0 \leq t \leq \tau\}$ which registers exactly the time-reversed jumps record. Finally the second measurement is performed using projectors $\{ \Theta \Pi_n^0 \Theta^\dagger\}$ which gives the (inverted) initial state $\Theta \ket{n}_0$. 

We will denote the complete time-reversed measurement record as $\tilde{\Gamma}_{[0,\tau]} =\{ m, (\tau - t_J, {k}_J),...,(\tau - t_1, {k}_1), n \}$, which reproduce the inverse sequence of (inverted) jumps with respect to $j_{[0,\tau]}$~\cite{Manzano22}. The trajectory operator associated to the time-reversed trajectory is:
\begin{equation}
\begin{split}
 \tilde{\mathcal{T}}({j}_{[0,\tau]}) =&~ \tilde{\mathcal{U}}(\tau,\tau -t_1)\tilde{L}_{k_1}~... \\ &~\tilde{\mathcal{U}}(\tau - t_{J-1},\tau - t_J)\tilde{L}_{k_J}\tilde{\mathcal{U}}(\tau-t_J,0),
\end{split}
 \end{equation}
with corresponding time-reversed operators for no-jumps periods and jumps: 
\begin{align}
 \tilde{\mathcal{U}}(t_2, t_1) &= \Theta~ T_+ e^{i\int_{t_1}^{t_2}dt(H(\tilde{\lambda}) - i\sum L_k^{\dagger}L_k)/2}~ \Theta^\dagger, \\
\tilde{L}_{k_j} &= \Theta L_{k_j}^\dagger \Theta^\dagger e^{-\Delta s_k/2} = \Theta  L_{\tilde{k}_j} \Theta^\dagger.
\end{align}
Notice that in the operator for no-jump periods in the time-reversed dynamics the Hamiltonian is now evaluated at value $\tilde{\lambda}(t) = \lambda(\tau - t)$, as it corresponds to the time-reversed driving protocol $\tilde{\Lambda}$. Moreover we notice that the jumps in the time-reversed dynamics correspond to the (inverted) complementary jumps $L_{\tilde{k}_j}$ by virtue of the detailed balance relation for operators $L_{\tilde k} = L_k^\dagger e^{-\Delta s_k/2}$, and $\Delta s_k$ is the entropy change in the environment associated to the kth jump~\cite{Manzano2018b,Manzano22}. 

The path probability for the inverse trajectory then reads:
\begin{equation}
    \tilde{\mathbb{P}}(\tilde{j}_{[0,\tau]}) = p_{m}(\tau)~ \tr \left[ \Theta \Pi_{n}^0 \Theta^\dagger~ \tilde{\mathcal{T}}({j}_{[0,\tau]})~ \Theta \Pi_{m}^\tau \Theta^\dagger~ \tilde{\mathcal{T}}^{\dagger}({j}_{[0,\tau]})\right].
    \label{eq:completeprobtr}
\end{equation}

Comparing the probability of trajectories in the forward and time-reversed processes we recover the link between irreversibility as measured in theoretic-information terms with thermodynamic entropy production as given in Eq.~\eqref{eq:Stot}, that is
\begin{eqnarray} \label{eqs:Stot}
    S_\mathrm{tot}(\tau) = \ln \left( \frac{\mathbb{P}(\Gamma_{[0,\tau]})}{\tilde{\mathbb{P}}(\tilde{\Gamma}_{[0,\tau]})} \right) = \Delta S_\mathrm{sys}(\tau) + \sum_r \beta_r Q_r(\tau),~~~
\end{eqnarray}
where the second equality follows by splitting the entropy production in the stochastic change in system's entropy (self-information) $\Delta S_\mathrm{sys}(\tau) = -\ln p_m(\tau) + \ln p_n(0)$ and the entropy exchange with the environment $\sum_r \beta_r Q_r(\tau) = \sum_j \Delta s_{k_j}$.  Assuming thermal reservoirs we obtain $\Delta s_k = \beta_r \Delta e_r$ with $\Delta e_r$ the energy transferred to the reservoir $r$ in the jump $k \in \mathcal{K}_r$ pertaining to the set of jumps induced by that reservoir.

\section{Information-theoretical irreversibility estimator from visible trajectories} \label{app5}

In this appendix we show how the estimator of irreversibility along visible trajectories  $\Sigma(\gamma_{[0,\tau]})$ can be obtained by explicitly calculating the probability of visible trajectories in the forward and time-reversed dynamics. In the case of imperfect detectors leading to hidden jumps during the evolution, the state of the system conditioned to a given measurement record $\gamma_{[0,\tau]}$ can be obtained from the stochastic master equation~(5) in the main text, which follows by a coarse-graining of outcomes in the ideal measurement scheme constructed for infinitesimal time-steps of the evolution. 

When a (visible) jump is detected, the evolution of the imperfectly monitored system state becomes, up to a normalization factor:
\begin{equation}
\begin{split}
\bar{\sigma}_{t+dt} &=  \eta_k M_k \rho M_k^\dagger \\ &= { \eta_k} L_k {\sigma_{t}}L_k^{\dagger} dt =: \mathcal{J}_k( \sqrt{\eta}_k {\sigma_t}) dt,    
\end{split}
\end{equation}
where we used the measurement operators for a jump $M_k = \sqrt{dt} L_k$ as in the ideal case, and used an overbar over the state, $\bar{\sigma}$, to remark that this state is not normalized. On the other hand the evolution during a step of time in which no jumps are detected is now given by a statistical combination of the no-jump evolution and the possibility that a hidden jump occurs, that is
\begin{equation}
\begin{split}
  \bar{\sigma}_{t+dt} &= M_0 \sigma_t M_0^\dagger + \sum_k (1- \eta_k)M_k \sigma_t M_k^\dagger \\ &=: {\sigma}_t + dt\mathcal{L_\ast}({\sigma}_t) + O(dt^2), 
\end{split}
\end{equation}
where we have introduced the superoperator: 
\begin{equation}
 \begin{split}
\mathcal{L}_\ast(\sigma) :=& -i\left[H,\sigma\right] -\frac{1}{2}\sum_k \left\{ L_k^{\dagger}L_k, \sigma \right\} \\ &+ \sum_k (1-\eta_k) L_k\sigma L_k^{\dagger}, 
\end{split}
\end{equation}
which includes the contributions of both no-jumps (second term) and hidden jumps (third term).

Given a visible measurement record $v_{[0,\tau]} = \{(t'_1, k'_1), ..., (t'_V, k'_V) \}$ with $V \leq J$ and some arbitrary initial state $\sigma_0$, one can recover the (unnormalized) state by sequential application of the above two evolution operators:
\begin{equation}
    \bar{\sigma} (\tau) = \mathcal{N}_{\tau,t_V} \circ \mathcal{J}_{k_V} \circ ~...~ \circ \mathcal{J}_{k_1} \circ \mathcal{N}_{t_1,0}(\sigma_0),
\end{equation}
where $\mathcal{N}_{t_2, t_1}(\sigma) := T_+ \exp[\int_{t_1}^{t_2} dt \mathcal{L}_\ast](\sigma)$ and we denoted by $A \circ B (\sigma) = A(B(\sigma))$ the convolution of superoperators. Notice that in this case we cannot define trajectory operators as in the ideal case. This is a manifestation of the breakdown of microscopic reversibility within the trajectories~\cite{Manzano22}.

The complete path probability of a visible trajectory given the measurement record $\gamma_{[0,\tau]}$ (including initial and final projective measurement outcomes) is then:
\begin{equation}\label{eqs:pathprobforward}
\begin{split}
    \mathrm{P}(\gamma_{[0,\tau]}) = p_n(0) ~\tr &\Big[ \Pi_{m}^\tau \mathcal{N}_{\tau,t_V} \circ \mathcal{J}_{k_V} \circ... \\ &... \circ \mathcal{J}_{k_1} \circ \mathcal{N}_{t_1,0}(\Pi_n^0) \Big],
\end{split}
\end{equation}
to be compared with the ideal case in Eq.~\eqref{eqs:completeprob}. Analogously to the ideal case, the probability of the time-reversed visible trajectory is given by:
\begin{equation}\label{eqs:pathprobbackward}
\begin{split}
    \tilde{\mathrm{P}}(\tilde{\gamma}_{[0,\tau]}) = p_m(\tau) &~ \tr  \Big[ \Theta \Pi_{n}^0 \Theta^\dagger \tilde{\mathcal{N}}_{\tau,\tau - t_1} \circ \tilde{\mathcal{J}}_{k_1} \circ... \\ &...\circ \tilde{\mathcal{J}}_{k_V} \circ \tilde{\mathcal{N}}_{\tau-t_V,0}(\Theta \Pi_m^\tau \Theta^\dagger) \Big],
\end{split}
\end{equation}
where we used the time-reversed versions of the superoperators in both jump intervals and smooth evolution periods:
\begin{align} \label{eq:bjumps}
  \tilde{\mathcal{J}}_k({\sigma}) :=& \eta_k \tilde{L}_k \sigma \tilde{L}_k^\dagger = { \eta_k} (\Theta L_k^{\dagger} \Theta^\dagger) {\sigma} (\Theta L_k \Theta^\dagger) e^{-\Delta s_k} \nonumber \\ =& \eta_k (\Theta {L}_{\tilde k} \Theta^\dagger) \sigma (\Theta {L}_{\tilde k}^\dagger \Theta^\dagger) \\
  \tilde{\mathcal{N}}_{t_2, t_1}(\sigma) :=& \Theta T_+ \exp[\int_{t_1}^{t_2} dt \tilde{\mathcal{L}}_\ast](\sigma) \Theta^\dagger
 \end{align}
with the time-reversal version of the superoperator $\mathcal{L}_\ast$ being:
\begin{equation} \label{eq:bnojumps}
 \begin{split}
 \tilde{\mathcal{L_\ast}}(\sigma) =& - i\left[H(\tilde{\lambda}),\sigma\right] -\frac{1}{2}\sum_k \left\{ L_k^{\dagger}L_k, \sigma \right\} \\ &+ \sum_k (1-\eta_k) {L}_{\tilde k}^{~} \sigma {L}_{\tilde k}^\dagger.
 \end{split}
\end{equation}
We note that here as before the Hamiltonian should be evaluated at the time-reversed value of the control parameter $\Lambda$, that is, $\tilde{\lambda}(t) = \lambda(\tau - t)$, and the inverse jumps ${L}_{\tilde k}$ appear in the third term (the second term is invariant with respect to the change $L_k \rightarrow {L}_{\tilde k}$) associated to efficiency $\eta_k$.

Importantly, in the above Eqs.~\eqref{eq:bjumps}-\eqref{eq:bnojumps} each type of jump (or  channel) appears associated to the efficiency of the inverse jump (or complementary channel) both for visible and hidden processes. That is, in the time-reversed imperfectly monitored process, the (inverse) jumps produced by operators $L_{\tilde k}$ are associated to an efficiency $\eta_k$, not to $\eta_{\tilde k}$, as in the original (forward) process. This is a consequence of applying the definitions for the time-reversed jumps in the ideal scenario, together with local detailed balance, for both visible and hidden jumps:
\begin{align}
 \tilde{L}^\prime_k &= \Theta {L}^{\prime~\dagger}_k \Theta^\dagger e^{-\Delta s_k/2}  = \sqrt{\eta_k} \Theta L^\dagger_k \Theta^\dagger e^{-\Delta s_k/2} \nonumber \\ 
 &= \sqrt{\eta_k} \Theta L_{\tilde k} \Theta^\dagger = \sqrt{\eta_k} \tilde{L}_k,~\\
  \tilde{L}^\ast_k &= \Theta {L}^{\ast~\dagger}_k \Theta^\dagger e^{-\Delta s_k/2} = \sqrt{1 - \eta_k} \Theta L^\dagger_k \Theta^\dagger e^{-\Delta s_k/2}  \nonumber \\ &= \sqrt{1 - \eta_k} \Theta L_{\tilde k} \Theta^\dagger = \sqrt{1 - \eta_k} \tilde{L}_k.
\end{align}
Therefore the actual implementation of the time-reversed process for imperfect monitoring requires not only an inversion of the driving protocol from $\Lambda$, but also implies an exchange in the efficiencies between complementary channels (e.g. emission efficiency should be exchanged with absorption efficiency, etc). The above exchange of efficiencies ensures that even if some of the channels coming in pairs are not directly monitored (e.g. $\eta_k > 0$ but $\eta_{\tilde k} = 0$), the estimator $\Sigma(\tau)$ introduced in this paper would still give us a lower bound on the underlying entropy production, avoiding any (spurious) divergence.

By comparing the probabilities of a given visible trajectory $\gamma_{[0,\tau]}$ in Eq.~\eqref{eqs:pathprobforward} with its inverse in Eq.~\eqref{eqs:pathprobbackward}, we obtain the irreversibility estimator along stochastic trajectories with imperfect monitoring as: 
\begin{equation} \label{eqs:Sigma}
\Sigma(\tau) = \mathrm{ln} \left( \frac{\mathrm{P}(\gamma_{[0,\tau]})}{\tilde{\mathrm{P}}(\tilde{\gamma}_{[0,\tau]})} \right) = \Delta S_\mathrm{sys}(\tau) + \phi(\tau), 
\end{equation}
where in the second equality we split the entropy production into system and environmental contributions with:
\begin{eqnarray} \label{eqs:Senv}
     &\phi(\tau) = \ln \left( \frac{\mathrm{P}(\gamma_{[0,\tau]}| n)}{\tilde{\mathrm{P}}(\tilde{\gamma}_{[0,\tau]}|m)} \right) \\ 
     &=\ln \left( \frac{\tr \left[ \Pi_{m}^\prime \mathcal{N}_{\tau,t_V} \circ \mathcal{J}_{k_V} \circ... \circ \mathcal{J}_{k_1} \circ \mathcal{N}_{t_1,0}(\Pi_n) \right]}{\tr \left[ \Theta \Pi_{n} \Theta^\dagger \tilde{\mathcal{N}}_{\tau,\tau - t_1} \circ \tilde{\mathcal{J}}_{k_1} \circ...\circ \tilde{\mathcal{J}}_{k_V} \circ \tilde{\mathcal{N}}_{\tau-t_V,0}(\Theta \Pi_m \Theta^\dagger) \right]} \right). \nonumber
\end{eqnarray}

From the expression in Eq.~\eqref{eqs:Sigma} it immediately follows, by taking the average over visible trajectories $\gamma_{[0,\tau]}$:
\begin{equation}
 \begin{split}
    \langle \Sigma(\tau) \rangle &= \sum_{\gamma_{[0,\tau]}} P(\gamma_{[0,\tau]}) \mathrm{ln} \left( \frac{\mathrm{P}(\gamma_{[0,\tau]})}{\tilde{\mathrm{P}}(\tilde{\gamma}_{[0,\tau]})} \right) \\ &= D \left[\mathrm{P}(\gamma_{[0,\tau]}) || \tilde{\mathrm{P}}(\tilde{\gamma}_{[0,\tau]})\right] \geq 0,
\end{split}
\end{equation}
where the last inequality follows from the fact that the Kullback-Leibler divergence is non-negative and zero if and only if $\mathrm{P}(\gamma_{[0,\tau]}) = \tilde{\mathrm{P}}(\tilde{\gamma}_{[0,\tau]})$. Analogously, from Eq.~\eqref{eqs:Sigma} it also follows an integral fluctuation theorem for $\Sigma$ as:
\begin{equation}
 \begin{split}
    \langle e^{- \Sigma} \rangle &= \sum_{\gamma_{[0,\tau]}} P(\gamma_{[0,\tau]}) e^{- \Sigma} \\ &= \sum_{\gamma_{[0,\tau]}} \tilde{P}(\tilde{\gamma}_{[0,\tau]}) = 1,
\end{split}
\end{equation}
which is valid whenever initial and final density operators share support.

\section{Proof of the main fluctuation theorem and inequalities} \label{app2}
In the following we provide a proof of the main fluctuation theorem presented in Eq.~\eqref{eq:FT}:
\begin{align}
 \langle e^{-S_\mathrm{tot}(\tau)} | \gamma_{[0,\tau]} \rangle &=  \sum_{h_{[0,\tau]}} \mathrm{P}(h_{[0,\tau]} | \gamma_{[0,\tau]}) \frac{\tilde{\mathbb{P}}(\tilde{\Gamma}_{[0,\tau]})}{\mathbb{P}(\Gamma_{[0,\tau]})} \\
 &= \sum_{h_{[0,\tau]}} \frac{\tilde{\mathbb{P}}(\tilde{\Gamma}_{[0,\tau]})}{\mathrm{P}(\gamma_{[0,\tau]})} = \frac{\tilde{\mathrm{P}} (\tilde{\gamma}_{[0,\tau]})}{\mathrm{P}(\gamma_{[0,\tau]})} = e^{-\Sigma(\tau)}, \nonumber
 \end{align}
where in the second line we used $\mathrm{P}(h_{[0,\tau]} | \gamma_{[0,\tau]}) = \mathbb{P}(\Gamma_{[0,\tau]})/ \mathrm{P}(\gamma_{[0,\tau]})$, then we performed the marginalization over sequences of hidden jumps $\sum_{h_{[0,\tau]}} \tilde{\mathbb{P}}(\tilde{\Gamma}_{[0,\tau]}) = \tilde{\mathrm{P}}(\tilde{\gamma}_{[0,\tau]})$, and in the last equality we identified the irreversibility indicator $\Sigma(\tau) = \log[ \mathrm{P}(\gamma_{[0,\tau]})/\tilde{\mathrm{P}}(\tilde{\gamma}_{[0,\tau]})]$ as defined in Eq.~\eqref{eq:Sigma}. 

We remark that here above the probability of visible trajectories in the time-reversed process $\tilde{P}(\tilde{\gamma}_{[0,\tau]})$ is implicitly defined by marginalization of $\tilde{\mathbb{P}}(\tilde{\Gamma}_{[0,\tau]})$ over the hidden sequences of jumps as defined from the set $\{L_j^\ast \}$ in the original (forward) process. This implies an effective exchange of the efficiencies of pairs of jumps in the time-reversed process, e.g. the efficiency of emissions in the time-reversed process becomes the efficiency of absorptions and viceversa (see App.\ref{app3} below for more details).

We now apply Jensen's inequality for conditional averages, namely, $f(\langle X | \gamma_{[0,\tau]}) \leq \langle f(X) | \gamma_{[0,\tau]} \rangle$ for any convex function $f(x)$ and trajectory functional $X(\Gamma_{[0,\tau]})$. For the case $f(x) = e^{-x}$ with $x = S_\mathrm{tot}- \Sigma$, we obtain:
\begin{equation}
    e^{-\langle S_\mathrm{tot} - \Sigma| \gamma_{[0,\tau]} \rangle} \leq \langle e^{-S_\mathrm{tot} + \Sigma} | \gamma_{[0,\tau]} \rangle = 1,
\end{equation}
which implies $\langle S_\mathrm{tot} | \gamma_{[0,\tau]} \rangle  - \Sigma \geq 0$ by just taking logarithms on both sides of the inequality and multiplying them by $-1$.
Finally, taking the average over all visible trajectories $\gamma_{[0,\tau]}$ we further obtain the  inequality:
\begin{eqnarray}
 \langle S_\mathrm{tot} \rangle  \geq  \langle \Sigma \rangle \geq 0.  
\end{eqnarray}
The above inequality can be interpreted as a stronger version of the second-law inequality $\langle S_\mathrm{tot} \rangle \geq 0$. Here $\langle \Sigma \rangle \geq 0$ follows from the fact that $\langle \Sigma \rangle$ can be written as a Kullback-Leibler divergence for the path probabilities of visible trajectories, see Sec.~\ref{sec:3} and App.~\ref{app5} below.

Finally, the above inequality can be extended to higher moments of $S_\mathrm{tot}$ and $\Sigma$. This follows by applying Jensen's inequality for conditional averages with $f(x) = [- \ln(x)]^k$, which is convex for even powers $k = 2, 4, 6, ...$ and take $x= e^{-S_\mathrm{tot}}$. This yields:
\begin{equation}
\begin{split}
\langle S_\mathrm{tot}^k | \gamma_{[0,\tau]} \rangle &\geq [-\ln \langle e^{-S_\mathrm{tot}} | \gamma_{[0,\tau]} \rangle]^k \\ 
&= [- \ln e^{- \Sigma}]^k = \Sigma^k,
\end{split}
\end{equation}
where we have used the main fluctuation theorem [Eq.~\eqref{eq:FT}~]. This in turn implies, by taking averages over visible trajectories $\gamma_{[0,\tau]}$ that:
\begin{eqnarray}
     \langle S_\mathrm{tot}^k \rangle  \geq  \langle \Sigma^k \rangle \geq 0,  
\end{eqnarray}
for all even $k$. This relation ensures that not only the average, but all even moments of the total entropy production distribution are lower-bounded by those of the estimator $\Sigma$.

\section{Proof of statistical bounds on entropy production estimation.} \label{app3}
Here we provide the proofs for our statistical bounds for the maximum and minimum of the differences $S_\mathrm{tot} - \Sigma$ reported in Eqs.~(8) and (9) of the main text. 
To prove these inequalities, we first introduce the conditional probability distribution for entropy production, given a set of visible jumps $\gamma_{[0,\tau]}$ as:
\begin{equation}
 \begin{split}
    \mathrm{Pr}(S_\mathrm{tot}|\gamma_{[0,\tau]}) :=& \int_{-\infty}^{\infty} \sum_{h_{[0,\tau]}} P(h_{[0,\tau]} | \gamma_{[0,\tau]}) \\ &~\delta \left( S_\mathrm{tot} -\ln \frac{\mathbb{P}(\Gamma_{[0,\tau]})}{\tilde{\mathbb{P}}(\tilde{\Gamma}_{[0,\tau]})} \right) dS_\mathrm{tot},   
 \end{split}
\end{equation}
where $\delta(x)$ denotes the Dirac delta function. The first inequality [Eq.~(8)] lower bounding deviations of the average entropy production conditioned on a measurement record from the estimator $\Sigma$ is obtained by closely following the derivations in Refs.~\cite{Jarzynski08,Jarzynski2011} as follows:
\begin{align}
     &\mathrm{Pr}(S_\mathrm{tot} - \Sigma < - \xi) = \mathrm{Pr}(S_\mathrm{tot} < \Sigma - \xi) \nonumber \\ 
     &= \int_{-\infty}^{\Sigma -\xi} dS_\mathrm{tot}~ \mathrm{Pr}(S_\mathrm{tot}|\gamma_{[0,\tau]}) \nonumber \\ 
     &\leq \int_{-\infty}^{\Sigma -\xi} dS_\mathrm{tot}~ \mathrm{Pr}(S_\mathrm{tot}|\gamma_{[0,\tau]})~e^{-S_\mathrm{tot} + \Sigma - \xi} \nonumber \\
     &= e^{\Sigma- \xi} \int_{-\infty}^{\Sigma-\xi} dS_\mathrm{tot}~ \mathrm{Pr}(S_\mathrm{tot}|\gamma_{[0,\tau]}) e^{-S_\mathrm{tot}} \nonumber \\ 
     &\leq  e^{\Sigma- \xi} \int_{-\infty}^{\infty} dS_\mathrm{tot}~ \mathrm{Pr}(S_\mathrm{tot}|\gamma_{[0,\tau]}) e^{-S_\mathrm{tot}} \nonumber \\ &= e^{\Sigma- \xi} \langle e^{-S_\mathrm{tot}(\tau)} | \gamma_{[0,\tau]} \rangle = e^{-\xi},
\end{align}
where in the first inequality we used that $S_\mathrm{tot} < \Sigma - \xi$ inside the integral limits, and the fluctuation theorem Eq.(1) in the main text for the last equality, and we recall that here $\xi \geq 0$.

Analogously the proof of the upper bound in Eq.~(9) follows as:
\begin{align}
 &\mathrm{Pr}(S_\mathrm{tot} - \Sigma \geq \xi) = \mathrm{Pr}(S_\mathrm{tot} \geq  \Sigma + \xi) \nonumber \\
 &= \int_{\Sigma + \xi}^{\infty} dS_\mathrm{tot}~ \mathrm{Pr}(S_\mathrm{tot}|\gamma_{[0,\tau]}) \nonumber \\ 
     &\leq \int_{\Sigma + \xi}^{\infty} dS_\mathrm{tot}~ \mathrm{Pr}(S_\mathrm{tot}|\gamma_{[0,\tau]})~e^{q(S_\mathrm{tot} - \Sigma - \xi)} \nonumber \\
     &\leq e^{- q \xi} \int_{-\infty}^{\infty} dS_\mathrm{tot}~ \mathrm{Pr}(S_\mathrm{tot}|\gamma_{[0,\tau]}) e^{q (S_\mathrm{tot}-\Sigma)} \nonumber \\ 
     &=  e^{- q \xi} \langle e^{q(S_\mathrm{tot} - \Sigma)} | \gamma_{[0,\tau]} \rangle,
\end{align}
where we have used that $S_\mathrm{tot} \geq \Sigma + \xi$ inside the integration interval and hence $q S_\mathrm{tot} \geq q(\Sigma + \xi)$ for $q \geq 1$, after which we extended the integration interval to the left and identified the conditional average $\langle e^{q(S_\mathrm{tot} - \Sigma)} | \gamma_{[0,\tau]} \rangle$. Recall that again here we assumed $\xi \geq 0$.

From the application of Jensen's inequality (see App.~\ref{app2}) for the family of convex functions $f(x)=x^{-q}$ with $q \geq 1$ to the fluctuation theorem in Eq.~\eqref{eq:FT0}, we obtain that
\begin{eqnarray}
\langle e^{q(S_\mathrm{tot} - \Sigma)} | \gamma_{[0,\tau]} \rangle \geq 1,
\end{eqnarray}
which provides a set of inequalities ensuring that the upper bound in Eq.~(9) scales generally slower than an exponential.

From the large deviation principle~\cite{Touchette09} we obtain by taking the long time limit:
\begin{eqnarray}
    \lim_{\tau \rightarrow \infty} \langle e^{q \tau(\frac{S_\mathrm{tot} - \Sigma}{\tau})} | \gamma_{[0,\tau]} \rangle \simeq e^{\tau K(q)} 
\end{eqnarray}
 with the scaled cumulant generating function $K(q):=\lim_{\tau \rightarrow \infty} \ln \langle e^{q (S_\mathrm{tot}(\tau) - \Sigma)} | \gamma_{[0, \tau]} \rangle / \tau$. That allows us to minimize the upper bound in Eq.~(9) over $q$ in the long time limit: 
 \begin{align}
     \mathrm{Pr}(S_\mathrm{tot} - \Sigma \geq \xi) \leq \min_q e^{- q \xi} e^{\tau K(q)} = e^{- \xi} e^{\tau K(1)},
 \end{align}
where in the last equality we obtained $q=1$ from the fact that $K(q) \geq 0$ for $q\geq 1$ and that $K(q)$ is convex, as reported in the main text.

\section{Fluctuation theorem with average over final projective measurement } \label{app6}

In this appendix we provide extra versions of the FT in Eq.~\eqref{eq:FT}, only involving conditional heat contributions $Q_r(\tau)$ and the effective entropy flux $\phi(\tau)$, as well as versions containing further averages over final measurement outcomes, hence not depending on the TPM scheme.

To begin with, we recall that both the total entropy production $S_\mathrm{tot}(\tau)$ in Eq.~\eqref{eq:Stot} and the irreversibility estimator for imperfect monitoring $\Sigma(\tau)$ in Eq.~\eqref{eq:Sigma} can be split into two contributions for system and environment, being the system contribution $\Delta S_\mathrm{sys}(\tau) = - \ln p_m(\tau) + \ln p_n(0)$ equal in both equations, and only dependent on the initial and final measurement outcomes. This allows us to rewrite the FT in Eq.~\eqref{eq:FT} as:
\begin{align}
    &e^{-\Delta S_\mathrm{sys}(\tau) - \phi(\tau)} = \langle e^{-\Delta S_\mathrm{sys}(\tau) - \sum_r \beta_r Q_r(\Gamma_{[0,\tau]})} | \gamma_{[0,t]} \rangle \nonumber \\ &= e^{-\Delta S_\mathrm{sys}(\tau)}\langle  e^{-\sum_r \beta_r Q_r(\Gamma_{[0,\tau]})} | \gamma_{[0,t]} \rangle,
\end{align}
where in the second equality we have used that the average is conditioned on the initial and final measurement outcomes $n$ and $m$. Finally multiplying by $e^{\Delta S_\mathrm{sys}(\tau)}$ on both sides of the equality, we obtain:
\begin{equation} \label{eqs:FT3}
e^{-\phi(\tau)} = \langle  e^{-\sum_r \beta_r Q_r(\Gamma_{[0,\tau]})} | \gamma_{[0,t]} \rangle,
\end{equation}
which relates the heat dissipated into the environment with the effective entropy flux $\phi(\tau)$ estimated from the visible trajectories only.

In the following we perform an average over the final outcomes on both sides of Eq.~\eqref{eqs:FT3}. Recall that the dependence on $n$ and $m$ is contained in $\Gamma_{[0,\tau]}$ and $\gamma_{[0,\tau]}$, but not in $h_{[0,\tau]}$. We obtain:
\begin{equation}
 \begin{split}
&\sum_m P[m | v_{[0,\tau]},n) e^{-\phi(\tau)} \\ &=  \sum_m P[m | v_{[0,\tau]},n) \langle  e^{-\sum_r \beta_r Q_r(\Gamma_{[0,\tau]})} | \gamma_{[0,t]} \rangle \\ &= \langle  e^{-\sum_r \beta_r Q_r(\Gamma_{[0,\tau]})} | n, v_{[0,t]} \rangle,
 \end{split}
\end{equation}
where in the second equality he used that $\gamma_{[0,\tau]} = \{ n , v_{[0,\tau]}, m\}$. Similarly one can add an average over the initial measurement to obtain:
\begin{equation} \label{eq:FTnew}
\begin{split}
&\sum_{m, n} p_n(0) P[m | v_{[0,\tau]},n) e^{-\phi(\tau)}  \\ 
&= \langle  e^{-\sum_r \beta_r Q_r(\Gamma_{[0,\tau]})} | v_{[0,t]} \rangle, 
\end{split}
\end{equation}
which no longer depends on the initial and final projective measurements performed in the TPM scheme.

As a corollary of  Eq.~\eqref{eq:FTnew}, from Jensen's inequality we also obtain the associated inequality:
\begin{equation}
 \begin{split}
& \sum_r \beta_r \langle   Q_r(\Gamma_{[0,\tau]}) | v_{[0,t]} \rangle \\ 
&\geq    \sum_{m, n} p_n(0) P[m | v_{[0,\tau]},n) \phi(\tau).
 \end{split}
\end{equation}

\section{Conditional sampling of quantum Monte-Carlo trajectories} \label{app7}

The numerical method that we use to simulate the dynamics of the monitored system under imperfect detection turns out to be a variation of the original Quantum Monte Carlo~\cite{Wiseman2009}. We will first briefly discuss how the original method works and then we report how this method can be adapted for our purposes.

\begin{enumerate}
    
    \item A given initial pure state of the system $\ket{\psi}_0$ evolves according to the (non-unitary) dynamics given by
    \begin{equation}\label{sch}
        i\dfrac{d \ket{\psi}_t}{dt} = H_\mathrm{eff} \ket{\psi}_t
    \end{equation}
    where $H_\mathrm{eff} := H - i/2 \sum_{k} L_{k}^{\dagger}L_{k}$ is a non-Hermitian operator. For simplicity we here assume that $H$ and the $L_k$ do not depend on time. Disregarding terms up to order $\delta t^2$ in the evolution, we can express the state of the system after a duration of $\delta t$ as follows:
    \begin{equation}
        \ket{\psi}_{t + \delta t} = \left ( 1 - i H_\mathrm{eff} \delta t \right) \ket{\psi }_t. 
    \end{equation}
    Since $H_\mathrm{eff}$ is non-Hermitian, the new vector $\ket{\psi}_{t+\delta t}$ is not normalized in general. In  fact it can be shown that the norm after $\delta t$ is $N = 1 - \delta p $ where $\delta p =\sum_{k}\delta p_{k} =\sum_{k} \bra{\psi}_tL_{k}^{\dagger}L_{k} \ket{\psi}_t/\langle \psi | \psi \rangle_t$ is the total probability of a quantum jump of any type $k= 1, ..., K$ taking place in the interval $[t, t + \delta t]$.

    \item In light of the decrease of the wave-function norm with time in the above equation, a Quantum Monte Carlo method can be efficiently implemented as a Gillespie algorithm as follows. We choose a random number $\alpha$ between zero and one, which will represent the probability that no quantum jumps (of any type) occur during a given interval. Then we integrate Eq.~(\ref{sch}) above until a time $t_1$, such that its norm equals the generated number, $\braket{\psi | \psi}_{t_1} = \alpha$. At this time a jump occurs. Once we know the time $t_1$ at which the jump takes place, we use the (normalized) state of the system at that instant, $\ket{\psi}_{t_1}$, to calculate the probabilities of the different jumps $\delta p_{k}$ and, by trowing a second random number, we choose one of them according to their probabilities. Then the state of the system is updated according to the jump selected $\ket{\psi}_{t_1} \rightarrow L_k \ket{\psi}_{t_1}/\delta p_k$ and we repeat the procedure.
\end{enumerate}

As our interest focuses on obtaining conditional averages for a given visible trajectory $\gamma_{[0,\tau]}$, we need to adapt the above method to sample conditional trajectories, that is sequences $\Gamma_{[0,\tau]}$ compatible with $\gamma_{[0,\tau]}$ with their respective probabilities. To effectively achieve that, we partition the trajectory into distinct ``hidden" intervals. These intervals span between two visible jump detections, as well as the intervals from the initial time to the first visible jump and from the last visible jump to the final condition. These intervals are linked, as each one is connected to the next through a detected jump. The process for sampling the conditional trajectories necessitates the sampling over these interconnected ``hidden" intervals for a given sequence $\gamma_{[0,t]}$.

To construct the trajectory in a hidden interval preceding a detected jump, we start by establishing an initial state. This initial state can be the initial state $\ket{\psi}_0$ provided at $t=0$, or the state following a prior detected jump $\ket{\psi}_{\rm prior}$. Once this initial state is established, we proceed to simulate the system's dynamics using the Quantum Monte Carlo method, as elucidated in the preceding section, until we approach the time of the next visible jump.

Throughout this simulation, we deliberately enforce the occurrence of jumps exclusively from undetected channels. Upon reaching the precise instant of the next visible jump $t_{\rm next}$, we subject the simulated ``hidden" interval to an acceptance evaluation. The acceptance criteria encompass both the probability of any jump occurring at the precise instant $t_{\rm next}$, denoted as $\braket{\psi| \psi }_{t_{\rm next}}$, as well as the likelihood of the specific observed (visible) jump $k_{\rm next}$, quantified by $\delta p_{k_{\rm next}}/\delta p$. With the interval's acceptance confirmed, we advance to the next hidden interval for which we set the next initial condition, thereby initiating the repetition of this entire process, until the final time of the simulation $\tau$ is reached. For the last hidden interval the acceptance criteria depends only on the probability of the projective measurement performed at the end, $\langle \psi | \Pi_m^\tau | \psi \rangle_{\tau}$.

The above procedure greatly speeds up the simulation of conditional trajectories with respect to the ''brute force'' method consisting on filtering trajectories in the whole interval $[0,\tau]$. However, it neglects statistical correlations between the hidden intervals, which makes it less accurate. For the simulations performed on the illustrative example reported in Sec.~\ref{sec:6}, we obtain good results for trajectories with few visible jumps in the considered intervals. However when increasing the density of visible jumps we start to see some small deviations. {We emphasize that this is a defect of the method we employ to numerically sample the trajectories, not about the general results.}

\bibliography{refs}

\end{document}